\documentclass[12pt]{elsart}
\usepackage{epsfig}

\let\I\i
\def\i{\mathrm{i}}

\def\d{\mathrm{d}}
\def\half{{\textstyle{1\over2}}}
\def\thalf{{\textstyle{3\over2}}}
\def\h{{\scriptscriptstyle{1\over2}}}
\def\th{{\scriptscriptstyle{3\over2}}}

\let\oldvec\vec
\def\pol#1{\oldvec{#1}}
\def\vec#1{\mbox{\boldmath$#1$}}
\def\svec#1{\mbox{{\scriptsize \boldmath$#1$}}}

\begin{document}

\begin{frontmatter}

\title{Axial amplitudes for $\Delta$ excitation in chiral quark models}

\author[PeF,IJS,EM0]{B.~Golli},
\author[FMF,IJS,EM1]{S.~\v{S}irca},
\author[Covilha,CCP,EM2]{L.~Amoreira}, and
\author[Coimbra,CCP,EM3]{M.~Fiolhais}

\address[PeF]{Faculty of Education,
              University of Ljubljana,
              1000 Ljubljana, Slovenia}
\address[FMF]{Faculty of Mathematics and Physics,
              University of Ljubljana,
              1000 Ljubljana, Slovenia}
\address[IJS]{J.~Stefan Institute, 
              1000 Ljubljana, Slovenia}
\address[Covilha]{Department of Physics,
                  University of Beira Interior,
                  6201-001 Covilh\~a, Portugal}
\address[Coimbra]{Department of Physics,
                  University of Coimbra,
                  3004-516 Coimbra, Portugal}
\address[CCP]{Centre for Computational Physics, 
                    University of Coimbra, 
                    3004-516 Coimbra, Portugal}

\thanks[EM0]{E-mail: bojan.golli@ijs.si}
\thanks[EM1]{E-mail: simon.sirca@ijs.si}
\thanks[EM2]{E-mail: amoreira@dfisica.ubi.pt}
\thanks[EM3]{E-mail: tmanuel@teor.fis.uc.pt}

\date{\today}

\begin{abstract}
We study the axial amplitudes for the N-$\Delta$ transition 
in models with quarks and chiral mesons.  A set of constraints
on the pion field is imposed which enforces PCAC and the
off-diagonal Goldberger-Treiman relation.  The quark contribution
to the amplitudes in general strongly underestimates
the $C^\mathrm{A}_5$ amplitude as well as the $\pi\mathrm{N}\Delta$
strong coupling constant.  We show that the results are considerably
improved in models that, in addition to the pion cloud, incorporate
a fluctuating $\sigma$ field inside the baryon.
\end{abstract}


\end{frontmatter}


\newpage

\section{Introduction}

The structure of the weak axial N-$\Delta$ transition currents
is ideally probed in neutrino or charged-lepton scattering experiments
on deuterium or hydrogen.  The experimental efforts so far have been
focused on the determination of the dipole axial mass parameter
\cite{arnd99}, without an attempt to break down the transition
current into form-factors \cite{llewellyn72}.  Although a number
of phenomenological predictions for the dominant coupling
$C_5^\mathrm{A}(0)$ exist (see Table~I of \cite{mukh98}
for an exhaustive list), the dependence of the form-factors
on momentum transfer is very poorly known.  Data on the 
non-leading form-factors $C_3^\mathrm{A}(Q^2)$
and $C_4^\mathrm{A}(Q^2)$ are especially scarce \cite{barish79}.
New information on the weak axial form-factors is expected
from parity-violating electron scattering experiments planned
at Jefferson Laboratory \cite{pv}.

Theoretical investigation of axial transition amplitudes
in different versions of the quark model is of particular
interest since it may reveal the importance of non-quark
degrees of freedom in baryons, in particular the chiral mesons. 
Yet, except for the calculation in the non-relativistic quark model
\cite{Mukh}, there exist almost no model predictions for the axial
transition amplitudes.  This can be traced back to the difficulty
of incorporating consistently the pion field which is necessary
to describe the proper low-$Q^2$ behaviour of the amplitudes.

The lack of experimental and theoretical knowledge in the weak
sector is in contrast to the case of electro-excitation of the
$\Delta$ resonance, which has been extensively studied theoretically
in the constituent quark models \cite{warns90} as well as chiral
models \cite{wirzba87}, and experimentally \cite{electro_exp}.
In \cite{FGS} we have pointed out the important role played by
the pion cloud in the determination of electro-production
amplitudes, in particular to the E2/M1 and C2/M1 ratios.
This has later been confirmed in other chiral models \cite{Silva}
and dynamical approaches \cite{dynamical}.

The aim of this work is to study some general properties
of the axial amplitudes in chiral quark models and present
theoretical predictions in two typical representatives
of such models, the linear $\sigma$ model with quarks 
and the Cloudy Bag Model.  We derive a set of constraints
on the pion field which enforce the proper behaviour of
the amplitudes in the vicinity of the pion pole.
We also address the long-standing problem of a too low
$\pi\mathrm{N}\Delta$ coupling constant which rather
systematically appears in all quark models.  
Comparing the results in the two models we are able to draw
some general conclusions regarding the contribution of chiral 
mesons to the weak amplitudes as well as to the strong
$\pi\mathrm{N}\Delta$ form factor.

\section{The axial transition amplitude and the
off-diagonal Goldberger-Treiman relation}

The axial N-$\Delta$ transition amplitude is usually parameterized
in terms of the Adler form-factors $C^\mathrm{A}_i(Q^2)$ as
\footnote{
Definition of the transition current
with respect to the $\Delta^{++}$ brings an additional
isospin factor $\sqrt{3}$ to RHS of (\ref{Aexp}).}
\begin{eqnarray}
\langle \Delta^+(p')|A_{\alpha(a=0)}|\mathrm{N}^+(p)\rangle
 &=& \bar{u}_{\Delta\alpha}\,{C^\mathrm{A}_4(Q^2)\over M_\mathrm{N}^2}\,
     {p'}_\mu q^\mu u_\mathrm{N}
   - \bar{u}_{\Delta\mu}\,{C^\mathrm{A}_4(Q^2)\over M_\mathrm{N}^2}\,
     {p'}_\alpha q^\mu u_\mathrm{N}\nonumber\\
&&
   + \bar{u}_{\Delta\alpha}\,C^\mathrm{A}_5(Q^2)\,u_\mathrm{N}
   + \bar{u}_{\Delta\mu}\,{C^\mathrm{A}_6(Q^2)\over M_\mathrm{N}^2}\,
     q^\mu q_\alpha u_\mathrm{N}\>{,}
\label{Aexp}
\end{eqnarray}
where ${p'}_\mu=(M_\Delta;0,0,0)$,  ${u}_{\Delta\alpha}$
is the corresponding Rarita-Schwinger spinor,
$p$ is the four-momentum of the nucleon and $q^\mu=(k_0;0,0,k)$ is
the four-momentum of the incident weak boson.  Then
$k_0^2-k^2=q^2\equiv-Q^2$ and
$k_0 = (M_\Delta^2 - M_\mathrm{N}^2 -Q^2)/2 M_\Delta$.
For simplicity, we take the third isospin component ($a=0$)
of the axial current.  We have omitted from (\ref{Aexp})
the $C^\mathrm{A}_3(Q^2)$ term \cite{llewellyn72},
which is consistent with the prediction of quark models
in which quarks occupy only the $l=0$ state.

It is convenient to work with helicity amplitudes\footnote{
The helicity amplitudes are normally defined as the matrix
elements of the interaction Hamiltonian and contain an
additional factor $\sqrt{4\pi\alpha_W/2K_0}$,
e.~g.~$S^\mathrm{A} =\sqrt{4\pi\alpha_W/2K_0}\,\tilde{S}^\mathrm{A}$,
where $K_0=k_0(Q^2=0)$ and $\alpha_W$ is the weak fine-structure
constant.}
\begin{eqnarray}
  \tilde{S}^\mathrm{A} &=&
                  -\langle \Delta^+(p'),s_\Delta=\half\,|\,A^0_0(0)
                  \,|\,\mbox{p}(p),s_\mathrm{N}=\half\rangle\>{,}
\label{defS}\\
  \tilde{A}^\mathrm{A}_\th &=&
                  -\langle \Delta^+(p'),s_\Delta=\thalf\,|\,
                  \vec{\varepsilon}_+ \cdot \vec{A}^0(0)
                  \,|\,\mbox{p}(p),s_\mathrm{N}=\half\rangle\>{,}
\label{defA32}\\
  \tilde{A}^\mathrm{A}_\h &=&
                  -\langle \Delta^+(p'),s_\Delta=\half\,|\,
                  \vec{\varepsilon}_+ \cdot \vec{A}^0(0)
                  \,|\,\mbox{p}(p),s_\mathrm{N}=-\half\rangle\>{,}
\label{defA12}\\
  \tilde{L}^\mathrm{A} &=&
                  -\langle \Delta^+(p'),s_\Delta=\half\,|\,
                  \vec{\varepsilon}_0 \cdot \vec{A}^0(0)
                  \,|\,\mbox{p}(p),s_\mathrm{N}=\half\rangle\>{,}
\label{defL}
\end{eqnarray}
where $s$ denotes the third spin component, and $\vec{\varepsilon}$
are the usual polarisation vectors.  The helicity amplitudes
are related to the $C_i^\mathrm{A}$ form-factors by
\begin{eqnarray}
C_6^\mathrm{A}  & = & {M_\mathrm{N}^2\over k^2}\,
    \left[-\tilde{A}_\th^\mathrm{A}
  + \sqrt{3\over2}\tilde{L}^\mathrm{A}\right]\>{,}
\label{C6}\\
C_5^\mathrm{A}  & = &
  \sqrt{3\over2}\left({k_0\over k}\,\tilde{S}^\mathrm{A}
  - {k_0^2\over k^2}\,\tilde{L}^\mathrm{A}\right)
  + {k_0^2-k^2\over k^2}\,\tilde{A}_\th^\mathrm{A}\>{,}
\label{C5}\\
C_4^\mathrm{A}  & = &
 {M_\mathrm{N}^2\over kM_\Delta}\left[-\sqrt{3\over2}\,
   \tilde{S}^\mathrm{A}
 + {k_0k\over M_\mathrm{N}^2}\,C_6^\mathrm{A}  \right]\, .
\label{C4}
\end{eqnarray}
In the approximation with $C^\mathrm{A}_3=0$ we have only one
independent transverse amplitude, since in this case
$\tilde{A}^\mathrm{A}_\th=\sqrt{3}\tilde{A}^\mathrm{A}_\h$.

From (\ref{Aexp}) it follows that the divergence of the transition
axial current vanishes in the chiral limit
provided  $C^\mathrm{A}_6(Q^2) = M_\mathrm{N}^2 C^\mathrm{A}_5(Q^2)/Q^2$.
The pole behaviour of the $C^\mathrm{A}_6$ amplitude suggests
that it is related to the term in
the axial current responsible for the pion decay,
${A^\alpha_a}_{(\mathrm{pole})}(x) = f_\pi\partial^\alpha\pi_a(x)$,
where $f_\pi = 93\,\mathrm{MeV}$ is the pion decay constant.
Taking a finite mass for the pion the divergence does not vanish
but is replaced by PCAC:
\begin{equation}
 \langle \Delta^+(p')\,|\,\partial^\alpha A_{\alpha\,a}\,|\,
    \mathrm{N}^+(p)\rangle
  = -m_\pi^2\,f_\pi\langle \Delta^+(p')\,|\,\pi_a(0)\,|\,
    \mathrm{N}^+(p)\rangle\>{,}
\label{dA}
\end{equation}
where the transition matrix element of the pion field is related
to the strong form factor $G_{\pi\Delta\mathrm{N}}(Q^2)$ by 
\begin{equation}
  \langle \Delta^+(p')\,|\,\pi_0(0)\,|\,\mathrm{N}^+(p)\rangle
 = \i {G_{\pi\Delta N}(Q^2)\over 2M_\mathrm{N}}\,
 {\bar{u}_{\Delta\mu}\,q^\mu u_\mathrm{N}\over Q^2 + m_\pi^2}\,
 \sqrt{2\over3}\>{.}
\label{piDN}
\end{equation}
Assuming that ${A^\alpha_a}_{(\mathrm{pole})}(x)$ dominates
the $C^\mathrm{A}_6$ amplitude for $Q^2\rightarrow -m_\pi^2$,
we obtain the {\em off-diagonal Goldberger-Treiman relation\/}
\cite{llewellyn72,schreiner,Hemmert}:
\begin{equation}
  C^\mathrm{A}_5(Q^2) =
  f_\pi\,{G_{\pi\mathrm{N}\Delta}(Q^2)\over2M_\mathrm{N}}\,
  \sqrt{2\over3}\;,
  \qquad
  Q^2\rightarrow -m_\pi^2\>{.}
\label{GT0}
\end{equation}
For a smooth interpolating pion field we expect
that (\ref{GT0}) holds also for moderate $Q^2$
in the physically accessible region.

\section{Helicity amplitudes in chiral quark models}

For a variety of models involving quarks interacting with chiral
fields $\sigma$ and $\pol{\pi}$ the Hamiltonian can be written
in the form
\begin{equation}
   H = H^0_q + H_\sigma + \int\d \vec{r}\left\{\half\left[\pol{P}_\pi^2
        + (\nabla^2 + m_\pi^2)\pol{\pi}^2\right]
        + U(\sigma,\pol{\pi}) + \sum_a j_a\pi_a\right\}\>{,}
\label{Hchi}
\end{equation}
where $j_a$ is the quark source, $\pol{P}_\pi$ is the pion
conjugate momentum, $H^0_q$ and $H_\sigma$ are the free-quark
and the $\sigma$-meson terms, and $U(\sigma,\pol{\pi})$
is the meson self-interaction term.  
In the Cloudy Bag Model the $\sigma$ field
and the self-interaction term are absent, 
while in the linear $\sigma$-model all terms
are present and the self-interaction term is the Mexican-hat
potential (see (\ref{LSM}) below).  

Let $|\mathrm{N}\rangle$ and  $|\Delta\rangle$ be the exact
solution of the Hamiltonian for the ground state and for
the $\Delta$, respectively, with
$H|\mathrm{N}\rangle = E_\mathrm{N}|\mathrm{N}\rangle$ and
$H|\Delta\rangle=E_\Delta|\Delta\rangle$. 
Then $\langle \mathrm{N}|[H,\pol{P}_\pi]|\mathrm{N}\rangle
= \langle \Delta|[H,\pol{P}_\pi]|\Delta\rangle = 0$
and
$\langle \Delta|[H,\pol{P}_\pi]|\mathrm{N}\rangle=
\i(E_\Delta-E_\mathrm{N})^2\langle\Delta|\pol{\pi}|\mathrm{N}\rangle$.
Evaluating the commutators using (\ref{Hchi}) for $a=0$, we obtain
\begin{eqnarray}
 (-\Delta+m_\pi^2)\langle \mathrm{N}\,|\,\pi_0(\vec{r})\,|\,
  \mathrm{N}\rangle
  &=&
  -\langle \mathrm{N}\,|\,J_0(\vec{r})\,|\,\mathrm{N}\rangle\;,
\label{virN}\\
 (-\Delta+m_\pi^2)\langle\Delta\,|\,\pi_0(\vec{r})\,|\,
  \Delta\rangle
  &=&
  -\langle\Delta\,|\,J_0(\vec{r})\,|\,\Delta\rangle\;,
\label{virD}\\
 (-\Delta+m_\pi^2-(E_\Delta - E_\mathrm{N})^2)\langle\Delta\,|\,
  \pi_0(\vec{r})\,|\,\mathrm{N}\rangle
  & = &
  - \langle\Delta\,|\,J_0(\vec{r})\,|\,\mathrm{N}\rangle\;.
\label{virDN}
\end{eqnarray}
The sources on the RHS of (\ref{virN})-(\ref{virDN}) consist
of the quark term and the term originating from the meson
self-interaction (if present):
\begin{equation}
  J_0(\vec{r}) = j_0(\vec{r}) +
      {\partial U(\sigma,\pol{\pi})\over\partial\pi_0(\vec{r})}\>{.}
\label{Jt}
\end{equation}

These relations hold for the exact solutions of (\ref{Hchi}).
In an approximate computational scheme they can be used as
constraints.

We now show an important property of the axial transition amplitudes
between states which satisfy these virial relations.
We split the axial current into the non-pole and the pole part,
$\pol{A}^\alpha = \pol{A}^\alpha_{(\mathrm{non-pole})} 
+ \pol{A}^\alpha_{(\mathrm{pole})}$, 
where
\begin{eqnarray}
\pol{A}^\alpha_{(\mathrm{non-pole})} &=& 
    \bar{\psi}\gamma^\alpha\gamma_5\,\half\pol{\tau}\psi
  + (\sigma-f_\pi)\partial^\alpha{\pol{\pi}}
    - {\pol{\pi}}\partial^\alpha{\sigma}\>{,}\label{Anp}\\
\pol{A}^\alpha_{(\mathrm{pole})} &=&
 f_\pi\partial^\alpha \pol{\pi}\>{.}\label{Apole}
\end{eqnarray}
Since the pole part involves only the pion field we can
use (\ref{virDN}) to evaluate its contribution to the amplitudes.
Note that (\ref{virDN}) is equivalent to (\ref{piDN})
since in our model we can write the strong $\mathrm{N}$-$\Delta$
transition form-factor as
\begin{equation}
  {G_{\pi\mathrm{N}\Delta}(Q^2)\over 2M_\mathrm{N}} = {1\over\i k}
  \langle\Delta\,||J_0(0) ||\,\mathrm{N}\rangle\, .
\label{GpiND}
\end{equation}
We find $\tilde{A}^\mathrm{A}_{\th\,(\mathrm{pole})} = 0$ and
\begin{equation}
\qquad
  \tilde{S}_{(\mathrm{pole})}^\mathrm{A} = {k_0\over k}
  \tilde{L}_{(\mathrm{pole})}^\mathrm{A}
              =  {2\over3}\,
                 {G_{\pi \mathrm{N}\Delta}(Q^2)\over2M_\mathrm{N}}\;
                 {f_\pi\, k\, k_0\over Q^2+m_\pi^2} \, .
\label{Spole}
\end{equation}
The pole term (\ref{Apole}) contributes only to $C^\mathrm{A}_6$,
\begin{equation}
 C^\mathrm{A}_{6\,(\mathrm{pole})}
  = f_\pi\,{G_{\pi\mathrm{N}\Delta}(Q^2)\over 2M_\mathrm{N}}\,
    {M_\mathrm{N}^2\over Q^2 + m_\pi^2}\,\sqrt{{2\over3}}\>{,}
\label{C6pole}
\end{equation}
while $C^\mathrm{A}_{4\,(\mathrm{pole})}=
C^\mathrm{A}_{5\,(\mathrm{pole})}=0$.
We conclude that in models in which the pion contribution to
the axial current has the simple form $f_\pi\partial^\alpha\pi_a$
and the pion field satisfies the virial relation (\ref{virDN})
{\em there is no pion contribution to the $C^\mathrm{A}_4$ and 
$C^\mathrm{A}_5$ amplitudes\/} while $C^\mathrm{A}_6$
is almost entirely dominated by the pion pole.  In such models
only the quarks contribute to the $C^\mathrm{A}_4$ and
$C^\mathrm{A}_5$ amplitudes.   In this respect, the calculation
of $C^\mathrm{A}_5$ in a constituent quark model calculation
(e.~g.~\cite{Mukh}), is still legitimate.

\section{Constrained calculation in the linear $\sigma$-model}

The linear $\sigma$-model assumes the following form of $j_t$ and
${\mathcal U}$ \cite{BB}:
\begin{equation}
  j_t = \i\,g\sum_{i=1}^3\,\bar{q}_i\gamma_5\tau_t q_i\;,
\qquad
    {\mathcal U} = {\lambda\over4}\,
    ({\sigma}^2 + {\pol{\pi}}^2 - f_\pi^2)^2\; .
\label{LSM}
\end{equation}
Here $q_i$ is the quark bispinor for the valence orbit (assumed
to be different for the nucleon and the $\Delta$),
and $\lambda = (m_\sigma^2 - m_\pi^2)/2f_\pi^2$.  The free parameters
of the model are the coupling strength $g$ related to the
``constituent'' mass of the quark $gf_\pi$, and the mass
of the $\sigma$ meson $m_\sigma$.  
The model has been successfully applied to the description
of the nucleon and $\Delta$ properties.
So far the physical states have been constructed
from the mean-field solution using either cranking \cite{BC} 
or the Peierls-Yoccoz projection \cite{GR}.
In the latter method the mean-field solution for the pion
field is interpreted as a coherent state.

The mean-field solution fulfills the diagonal virial relations 
(\ref{virN})-(\ref{virD}) but not the off-diagonal
relation (\ref{virDN}).
To satisfy this relations it is necessary to include a
channel representing the $\Delta$ decay, i.e. a term
that asymptotically represents the nucleon and a free pion.
We have therefore taken a more general ansatz for the $\Delta$: 
\begin{equation}
  |\Delta\rangle = \mathcal{N}_\Delta\left\{
   P^\th\Phi_\Delta|\Delta_q\rangle +
   \int\d k\, \eta(k)
[a_{mt}^\dagger(k)|\mathrm{N}\rangle
]^{\th\th}\right\}\,,
\label{D2mode}
\end{equation}
where the first term represents the bare $\Delta$ state
surrounded by a cloud of pions and $\sigma$ mesons,
$P^\th$ is the projection operator on the subspace with 
isospin and   angular momentum $\thalf$,
$|\mathrm{N}\rangle$ is the nucleon ground state, and
$[\,\,\,]^{\th\th}$ denotes a pion-nucleon state with
isospin $\thalf$ and spin $\thalf$.
Requiring that the energy of this state is stationary, 
the denominator of $\eta(k)$ takes the form 
$\omega_k-(E_\Delta-E_\mathrm{N})$
which is also the form implied by (\ref{virDN}). 
For the nucleon we assume: 
\begin{equation}
  |\mathrm{N}\rangle = \mathcal{N}_\mathrm{N}
  P^\h\left[\Phi_\mathrm{N}|\mathrm{N}_q\rangle +
  \Phi_{\mathrm{N}\Delta}|\Delta_q\rangle\right] \>{.}
\label{N2mode}
\end{equation}
Here $\Phi_\mathrm{N}$ and $\Phi_{\mathrm{N}\Delta}$ stand 
for hedgehog coherent states describing the pion cloud around 
the bare nucleon and bare $\Delta$, respectively. 
To match the third constraint, (\ref{virDN}), the denominator
of the pion state in the second term of (\ref{N2mode}) 
should behave as $\omega_k+\omega_0$ with
$\omega_0=(E_\Delta-E_\mathrm{N})$.
In the above ansatz, only one profile for the $\sigma$ field 
is assumed\footnote{
Since the $\sigma$ field is scalar its analog of (\ref{virDN})
is identically zero.}.

The properties of the ground state are dominated by the first term
in (\ref{N2mode}), and imposing the off-diagonal constraint
influences only slightly the results.
For the $\Delta$, the inclusion of the decaying channel
modifies the long-range behaviour of the pion field,
and yields the correct low-$Q^2$ behaviour of the 
transition amplitudes as explained in the previous section.
The calculated $\Delta$-$\mathrm{N}$ splitting 
is typically only $(50-70)\,\%$ of the experimental value.  
In order to make a sensible comparison of the transition
amplitudes with the experimental ones, it is necessary
to have the correct kinematical relations in the model.
This can be achieved
by including an additional phenomenological term in the Hamiltonian
mimicking either the chromo-magnetic or the instanton-induced
interaction between quarks and adjusting its strength such as 
to bring the $\Delta$-$\mathrm{N}$ splitting 
to the experimental value.

\section{Calculation of the amplitudes}

We calculated the amplitudes in two models: in the linear $\sigma$-model
and in the Cloudy Bag Model.   In the Cloudy Bag Model we assume
the usual perturbative form for the pion profiles \cite{thomas}
using the experimental masses for the nucleon and $\Delta$,
which fulfills the virial constraints (\ref{virN})-(\ref{virDN}).
Since the pion contribution to the axial current 
in the Cloudy Bag Model has
the form of the pole term in (\ref{Apole}), only the quarks contribute
to the $C^\mathrm{A}_5$ and $C^\mathrm{A}_4$ amplitudes.

The amplitudes (\ref{defS})-(\ref{defL}) are defined between
states with good 4-momenta $p'$ and $p$ respectively while
in the model calculations localised states are used.  We can use
such states in our calculation of amplitudes by interpreting them
as wave packets of states with good linear momentum.  Extending
the method explained in \cite{Hemmert} we find for a chosen component
of the axial current evaluated between localised states,
$\langle\Delta|A(\vec{r})|\mathrm{N}\rangle$:
\begin{equation}
\int\d\vec{p}\,\varphi_\Delta^*(\vec{p}+\vec{k})\varphi_N(\vec{p})
  \langle\Delta(\vec{p}+\vec{k})| A(0)|\mathrm{N}(\vec{p})\rangle
 = \int\d\vec{r}\, e^{\i\svec{k}\svec{r}}
    \langle\Delta|A(\vec{r})|\mathrm{N}\rangle\>{.}
\label{M02Mr}
\end{equation}
Here the matrix element of (\ref{Aexp}) is taken on the LHS and
$\varphi_N(\vec{p})$ and $\varphi_\Delta(\vec{p})$ 
are (normalised) functions describing the center-of-mass
motion of the localised solution for the nucleon and the $\Delta$,
respectively.  We assume that the spread of the wave packet is
of the order of the inverse baryon mass ($M^{-1}$) and use
for simplicity the same spread for the nucleon and the delta.
The Adler form-factors of (\ref{C6})--(\ref{C4}) are then 
modified in such a way that $C_6^A$ and $C_5^A$ are multiplied
by the factor $2M_\Delta/(M_\Delta+M_\mathrm{N})$, while
\begin{equation}
C_4^\mathrm{A}  =
 {M_\mathrm{N}^2\over kM_\Delta}\left[-\sqrt{3\over2}\,\tilde{S}^\mathrm{A}
 + {k_0k\over M_\mathrm{N}^2}\,{M_\Delta+M_\mathrm{N}\over2M_\Delta}\,
   C_6^\mathrm{A}
   \right] - {M_N^2\over2M_\Delta^2}\,C_5^\mathrm{A} \>{.}
\label{C4wp}
\end{equation}
We have neglected terms of the order $k^2/M^2$.  Similarly,
the strong $G_{\pi\mathrm{N}\Delta}$ form factor (\ref{GpiND})
acquires the same correction factor.  The essential property
that the pole contribution cancels out in $C_4^\mathrm{A}$ and
$C_5^\mathrm{A}$ still persists as well as the relation (\ref{C6pole})
for $C_6^\mathrm{A}$.

\begin{figure}[ht]
\begin{center}
\includegraphics[height=9cm]{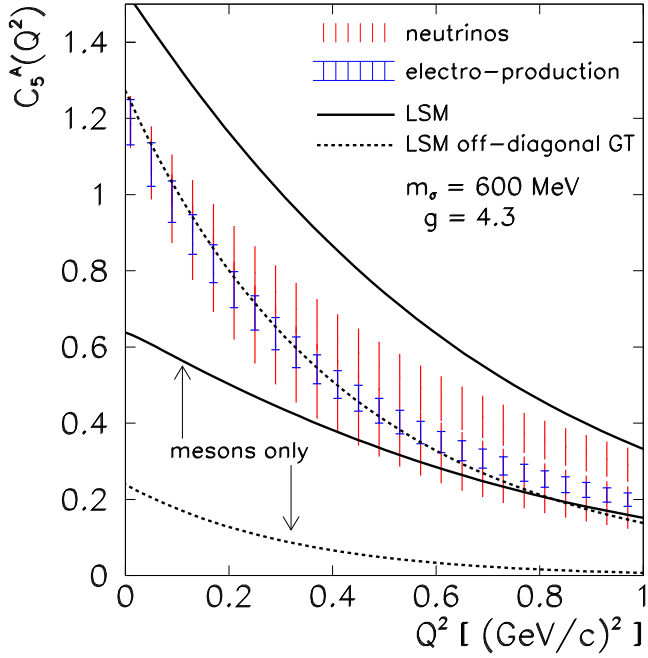}
\end{center}
\caption{The amplitude $C_5^\mathrm{A}(Q^2)$ in the 
linear $\sigma$-model.  The experimental value of $1.22\pm 0.06$
at $Q^2=0$ \protect\cite{ruso99} is based on data
from ANL and BNL \protect\cite{radecky82,kitagaki90}.
The error ranges are given by the spread in the axial-mass
parameter $M_\mathrm{A}$ as determined from neutrino scattering
experiments (broader range) and from electro-production of pions
\protect\cite{arnd99} ($M_\mathrm{A}=(1.077\pm 0.039)\,\mathrm{GeV}$,
narrower range).  Full curves: wave-packet result; dashed curves:
calculation from $G_{\pi\mathrm{N}\Delta}$ (\protect\ref{GT0}).}
\label{fig:c5a}
\end{figure}

Fig.~\ref{fig:c5a} shows the $C^\mathrm{A}_5$ amplitude 
in the linear $\sigma$-model
with $g=4.3$ and $m_\sigma=600\,\mathrm{MeV}$ compared to the
experimental weak axial form-factors given in the convention
of Adler \cite{adler68,adler75}, with a phenomenological dipole
parameterisation
$C_i^\mathrm{A}(Q^2) = C_i^\mathrm{A}(0) / (1 + Q^2/M_\mathrm{A}^2)^2$.
The $C_5^\mathrm{A}(0)$ is $25\,\%$ higher than the experimental
estimate, while the $M_\mathrm{A}$ from a dipole fit to our
calculated values matches the experimental
$M_\mathrm{A}$ to within a few percent.

We note that for the nucleon we obtain $g_\mathrm{A}=1.41$
which is roughly the same amount higher than the experimental
value of $1.27$.  On the other hand, if we determine
$C^\mathrm{A}_5(Q^2)$ from the calculated strong
$\pi\mathrm{N}\Delta$ form-factor using
the Goldberger-Treiman relation (\ref{GT0}) we obtain
a better agreement, yet with a steeper fall-off corresponding
to $M_\mathrm{A}\approx 0.80\,\mathrm{GeV}$.
The discrepancy between the two calculated values
($17\,\%$ at $Q^2=-m_\pi^2$ where (\ref{GT0}) holds) 
is a measure for the quality of our approximate computational
approach.  It can be attributed to a too large meson contribution
originating from the last two terms in (\ref{Anp}).
Since in this model only the meson
fields bind the quarks it is reasonable that their strength
is overestimated in the variational calculation.  The effect
of the meson self-interaction (the second term in (\ref{Jt})) is
relatively less pronounced in the strong coupling constant
(only $\sim 20\,\%$) than in $C^\mathrm{A}_5(Q^2)$.
Both $G_{\pi\mathrm{N}\Delta}(0)$ and $G_{\pi\mathrm{NN}}(0)$
are over-estimated in the model by $\sim 10\,\%$.  Still,
the ratio $G_{\pi\mathrm{N}\Delta}(0)/G_{\pi\mathrm{NN}}(0)=2.01$
is considerably higher than either the familiar SU(6)
prediction $\sqrt{72/25}$ or the mass-corrected value
of $1.65$~\cite{Hemmert}, and compares reasonably well
with the experimental value of $2.2$.  This improvement
is mostly a consequence of the renormalisation of the strong
vertices due to pions.

The value of $C^\mathrm{A}_5$ grows with $g$ and $m_\sigma$ in
contrast to $G_{\pi\mathrm{N}\Delta}$ which remains 
almost constant over a large range of model parameters.
In our calculation we cannot use much lower values for
$g$ since the solution becomes numerically unstable.

\begin{figure}[ht]
\begin{center}
\includegraphics[height=9cm]{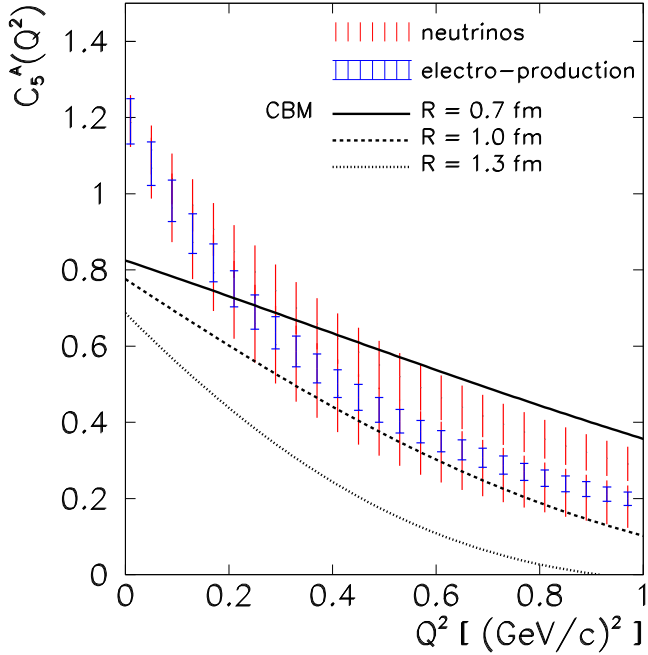}
\end{center}
\caption{The amplitude $C_5^\mathrm{A}(Q^2)$ in the Cloudy Bag Model
for three values of the bag radius.  Experimental uncertainties
are as in caption to Fig.~\protect\ref{fig:c5a}.}
\label{fig:c5a_CBM}
\end{figure}

In the Cloudy Bag Model the picture is reversed.  
Here only the first term
in (\ref{Anp}) contributes to the amplitudes; as a result the
$C^\mathrm{A}_5$ amplitude is less than $2/3$ of the
experimental value (see Fig.~\ref{fig:c5a_CBM}).
The behaviour of $C^\mathrm{A}_5$ is similar
as in the pure MIT Bag Model (to within $10\,\%$),
with fitted $M_\mathrm{A}\sim 1.2\,\mathrm{GeV\,fm}/R$.
The off-diagonal Goldberger-Treiman relation is satisfied
in the Cloudy Bag Model, but $C^\mathrm{A}_5$ 
from $G_{\pi\mathrm{N}\Delta}$
has a steeper fall-off with fitted
$M_\mathrm{A}\sim 0.8\,\mathrm{GeV\,fm}/R$.
The ratio $C^\mathrm{A}_5(0)/g_\mathrm{A}$ 
is close to the model-independent prediction of \cite{slaughter}.

The large discrepancy can be to some extent attributed to the
fact that the Cloudy Bag Model predicts a too low value for 
$G_{\pi\mathrm{NN}}$, and consequently $G_{\pi\mathrm{N}\Delta}$.
Taking a smaller value of $f_\pi$ in order to increase the
strong coupling constants does not improve the results since
$f_\pi$ on the RHS of (\ref{GT0}) compensates for the change
in $G_{\pi\mathrm{N}\Delta}$.  We have found that the pions
increase the $G_{\pi\mathrm{N}\Delta}/G_{\pi\mathrm{NN}}$ ratio
by $\sim 15\,\%$ through vertex renormalisation.
The effect is further enhanced by the mass-correction factor
$2M_\Delta/(M_\Delta+M_\mathrm{N})$, yet suppressed in
the kinematical extrapolation of $G_{\pi\mathrm{N}\Delta}(Q^2)$
to the $\mathrm{SU}(6)$ limit.  This suppression is weaker
at small bag radii $R$: the ratio drops from $2.05$
at $R=0.7\,\mathrm{fm}$ to $1.60$ (below the $\mathrm{SU}(6)$
value) at $R=1.3\,\mathrm{fm}$.

\begin{figure}[ht]
\begin{center}
\includegraphics[height=9cm]{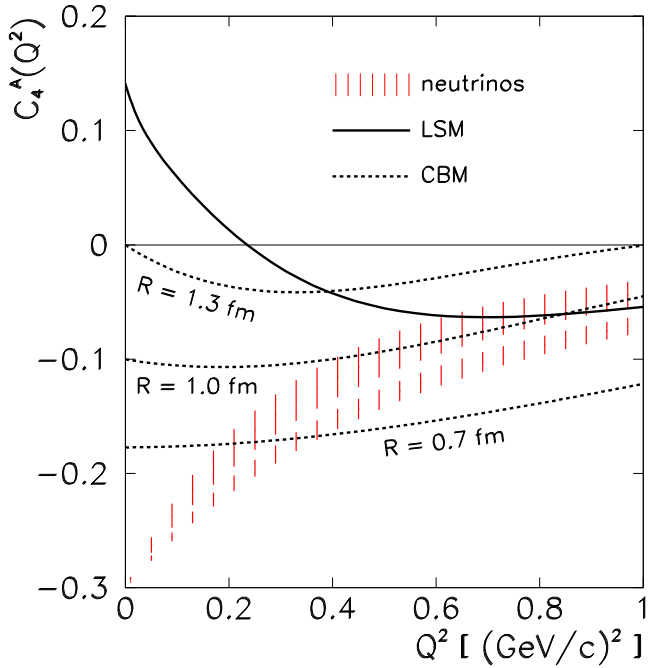}
\end{center}
\caption{The amplitude $C_4^\mathrm{A}(Q^2)$ in the 
linear $\sigma$-model, with model parameters and 
experimental uncertainties
due to the spread in $M_\mathrm{A}$ as in Fig.~\protect\ref{fig:c5a}.
Experimentally, $C^\mathrm{A}_4(0)=-0.3\pm 0.5$ \protect\cite{barish79}.
For orientation, the value for $C_4^\mathrm{A}(0)$ is used
without error-bars.}
\label{fig:c4a}
\end{figure}

The determination of the $C^\mathrm{A}_4$ is less reliable
because the meson contribution to the scalar amplitude
is very sensitive to small variations of the profiles.
However, the experimental value is very uncertain as well.
Neglecting the non-pole contribution to $S^\mathrm{A}$ and
$C_6^\mathrm{A}$ (the pole contribution cancels out) we see
from (\ref{C4wp}) that the value of $C_4^\mathrm{A}$ is dominated
by the term $-(M_\mathrm{N}^2/2M_\Delta^2)\,C_5^\mathrm{A}$,
in agreement with the popular parameterisation
of the amplitudes.  In our models, the non-pole contribution
to $C_6^\mathrm{A}$ is not negligible and tends to increase
$C_4^\mathrm{A}$ at small $Q^2$, as seen in Fig.~\ref{fig:c4a}.

\begin{figure}[ht]
\begin{center}
\includegraphics[height=9cm]{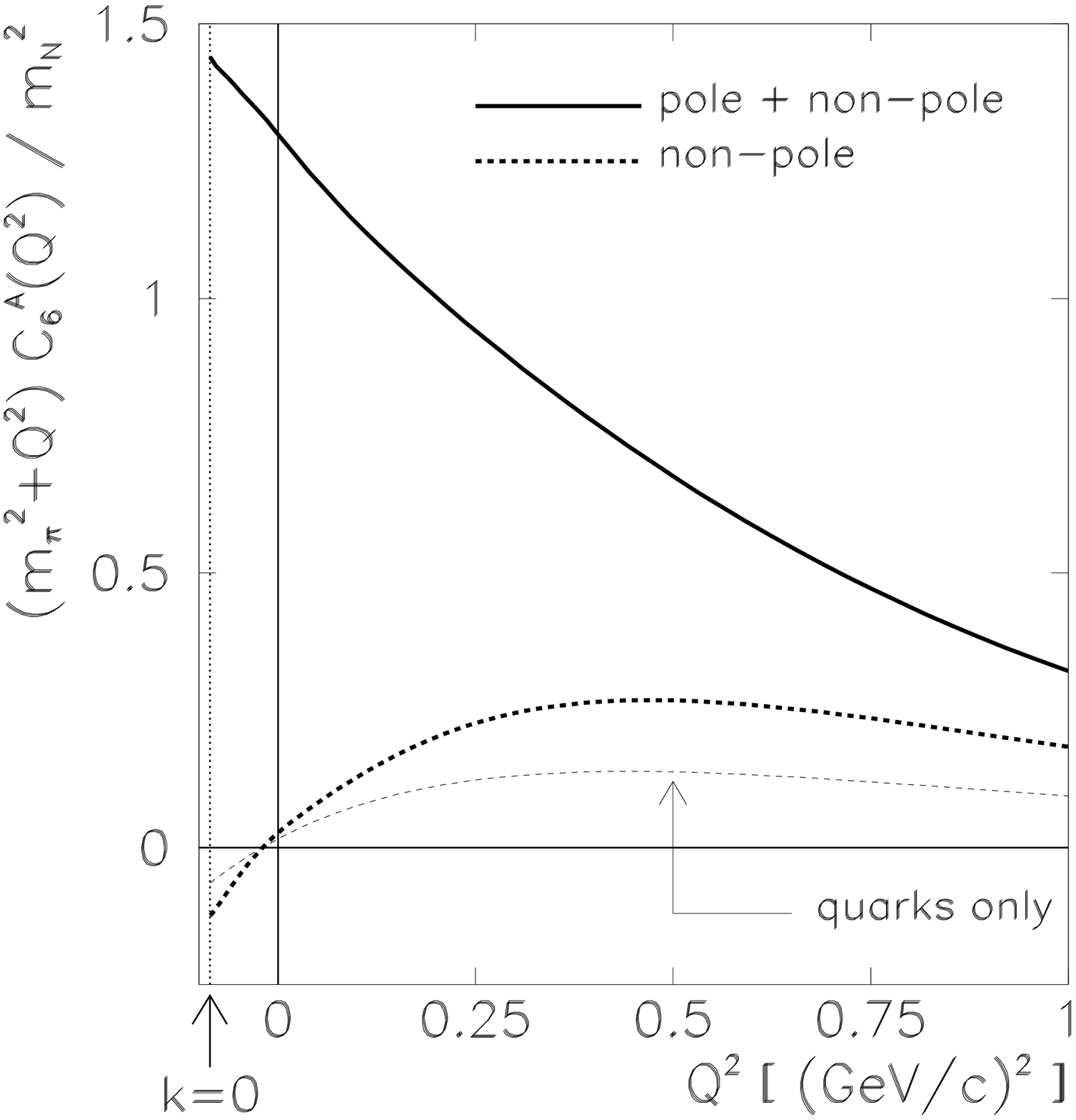}
\end{center}
\caption{The non-pole part and the total amplitude
$C_6^\mathrm{A}(Q^2)$ in the linear $\sigma$-model.
Model parameters are as in Fig.~\protect\ref{fig:c5a}.}
\label{fig:c6a}
\end{figure}

The $C^\mathrm{A}_6$ amplitude is governed by the pion pole
for small values of $Q^2$ and hence by the value
of $G_{\pi\mathrm{N}\Delta}$ which is well reproduced
in the linear $\sigma$-model, and underestimated by $\sim 35\,\%$ 
in the Cloudy Bag Model.
Fig.~\ref{fig:c6a} shows that the non-pole contribution becomes
relatively more important at larger values of $Q^2$.

\section{Conclusions}

To the best of our knowledge the present work is the first
attempt to calculate the axial $\mathrm{N}$-$\Delta$
transition amplitudes in a quark model which consistently 
includes the chiral mesons already at the Lagrangian level.
We have derived a set of constraints which ensures the
proper treatment of the pion pole dominating the transition
at low $Q^2$.  Though there is a rather strong discrepancy
between  calculated amplitudes in the two models considered
here, we are nonetheless able to draw some general conclusions 
about the role of the chiral mesons.

The quark contribution alone strongly underestimates
the $C_5^\mathrm{A}$ amplitude.  Models in which only a linear
coupling of pions to quarks is added do not improve the situation
since in such a case the pion term in the axial current does not
contribute to the amplitude.  On the other hand, the inclusion
of meson self-interaction which allows for a substantial deviation
of the $\sigma$ field from its vacuum value inside the baryon
considerably increases $C_5^\mathrm{A}$.  
The linear $\sigma$-model seemingly
overestimates this contribution as it could have been anticipated
from the overestimate of $g_\mathrm{A}$ obtained in this model.

Regarding the ratio $G_{\pi\mathrm{N}\Delta}/G_{\pi\mathrm{NN}}$ 
we find that it is the pion cloud which enhances its value
compared to the $\mathrm{SU}(6)$ value of $\sqrt{72/25}$;
in the linear $\sigma$-model as well as 
in the Cloudy Bag Model for smaller bag radii
the ratio is greater than $2$ and not far from 
the experimentally determined value of $2.2$.

The $Q^2$-behaviour of the axial amplitudes is well reproduced
in the linear $\sigma$-model.  We stress that the behaviour of  
$G_{\pi\mathrm{N}\Delta}(Q^2)$ is considerably softer,
with a cut-off parameter (corresponding to the axial 
mass $M_\mathrm{A}$) of $\sim 0.8\,\mathrm{GeV}$.
A similar trend is also seen in the Cloudy Bag Model for bag radii 
above $\sim 1\,\mathrm{fm}$.  The popular assumption
in which the same value for the strong and axial cut-offs
are taken is therefore not supported by the two models.

This work was supported by FCT (POCTI/FEDER), Lisbon, and by
The Ministry of Science and Education of Slovenia.


\begin{thebibliography}{99}
\bibitem{arnd99} 
A.~Liesenfeld et al., Phys.~Lett.~B {\bf 468} (1999) 20,
  and references therein.
\bibitem{llewellyn72} 
  C.~H.~Llewellyn Smith, Phys. Rep. {\bf 3C} (1972) 261.
\bibitem{mukh98} 
  N.~C.~Mukhopadhyay et al., Nucl. Phys. A {\bf 633} (1998) 481.
\bibitem{barish79} 
  S.~J.~Barish et al., Phys. Rev. D {\bf 19} (1979) 2521.
\bibitem{pv} L.~Elouadrhiri, D.~Heddle (spokespersons),
  JLab Proposal E94-005;\\
  N.~Simi\'cevi\'c, S.~P.~Wells (spokespersons),
  JLab Proposal PR97-104.
\bibitem{Mukh} 
  J\`un L\'\I u, N.~C.~Mukhopadhyay, L.~Zhang,
  Phys. Rev. C {\bf 52} (1995) 1630.
\bibitem{warns90} 
  M.~Warns et al., Z. Phys. Rev. C {\bf 45} (1990) 627;\\
  S. Capstick, B.~D. Keister, Phys. Rev. D {\bf 51} (1995) 3598;\\
  P.~Grabmayr, A.~J.~Buchmann, Phys. Rev. Lett. {\bf 86} (2001) 2237.
\bibitem{wirzba87} 
  A. Wirzba, W. Weise, Phys.~Lett.~B {\bf 188} (1987) 6;\\
  K. Bermuth et al., Phys. Rev. D {\bf 37} (1988) 89.
\bibitem{electro_exp} K.~Joo et al., Phys. Rev. Lett. {\bf 88} (2002) 122001;\\
  Th.~Pospischil et al., Phys. Rev. Lett. {\bf 86} (2001) 2959;\\
  C.~Mertz et al., Phys. Rev. Lett. {\bf 86} (2001) 2963.
\bibitem{FGS}
  M. Fiolhais, B. Golli, S.~\v Sirca,  Phys.~Lett.~B {\bf 373} (1996) 229.
\bibitem{Silva}
  D.~H.~Lu, A.~W.~Thomas, A.~G.~Williams, Phys. Rev. C {\bf 55} (1997) 3108;\\
  G.~C.~Gellas, T.~R.~Hemmert, C.~N.~Ktorides, G.~I.~Poulis,
  Phys. Rev. D {\bf 60} (1999) 054022;\\
  A. Silva et al., Nucl. Phys. A {\bf 675} (2000) 637;\\
  L.~Amoreira, P.~Alberto, M.~Fiolhais, Phys. Rev. C {\bf 62} (2000) 045202.
\bibitem{dynamical} 
  S.~S.~Kamalov, S.~N.~Yang, Phys. Rev. Lett. {\bf 83} (1999) 4494;\\
  T.~Sato, T.-S.~H.~Lee, Phys. Rev. C {\bf 63} (2001) 055201.
\bibitem{schreiner}
  P.~A.~Schreiner, F.~von~Hippel, Nucl. Phys. B {\bf 58} (1973) 333.
\bibitem{Hemmert}
  T.~R.~Hemmert, B.~R.~Holstein, N.~C.~Mukhopadhyay,
  Phys. Rev. D {\bf 51} (1995) 158.
\bibitem{BB} 
  M.~C. Birse, M.~K.~Banerjee, Phys. Lett. B {\bf 136} (1984) 284;\\
  M.~C. Birse, M.~K.~Banerjee, Phys. Rev. D {\bf 31} (1985) 118;\\
  S.~Kahana, G.~Ripka, V.~Soni, Nucl. Phys. A {\bf 415} (1984) 351.
\bibitem{BC}
 T.D. Cohen and W. Broniowski, Phys. Rev. D 34 (1986) 3472
\bibitem{GR}
  B.~Golli, M.~Rosina, Phys. Lett. B {\bf 165} (1985) 347;\\
  M.~C.~Birse, Phys. Rev. D {\bf 33} (1986) 1934.
\bibitem{thomas} 
  A.~W.~Thomas, Adv. Nucl. Phys. {\bf 13} (1984) 1;\\
  L. Amoreira et al., Int. J. Mod. Phys. {\bf 14} (1999) 731.
\bibitem{adler68} 
  S.~L.~Adler, Ann. Phys. {\bf 50} (1968) 89.
\bibitem{adler75} 
  S.~L.~Adler, Phys. Rev. D {\bf 12} (1975) 2644.
\bibitem{ruso99} 
L.~Alvarez-Ruso, S.~K.~Singh, M.~J.~Vicente-Vacas,
  Phys. Rev. C {\bf 59} (1999) 3386.
\bibitem{radecky82} 
  G.~M.~Radecky et al., Phys. Rev. D {\bf 25} (1982) 1161.
\bibitem{kitagaki90} 
T.~Kitagaki et al., Phys. Rev. D {\bf 42} (1990) 1331,
  and references therein.
\bibitem{slaughter} M.~D.~Slaughter, Nucl. Phys. A {\bf 703} (2002) 295.
\end{thebibliography}
\end{document}